\documentclass[a4paper,11pt]{article}
\usepackage{pos}

\title{Parton color reconnection in Herwig 7 using a spacetime event topology}

\author[a]{J. Bellm}
\author[b,c]{C. B. Duncan}
\author[c]{S. Gieseke}
\author*[d]{M. Myska}
\author[d,e]{A. K. Siodmok}

\affiliation[a]{Theoretical Particle Physics, Department of Astronomy and Theoretical Physics, Lund University\\ Lund, Sweden}

\affiliation[b]{School of Physics and Astronomy, Monash University\\
Clayton, VIC 3800, Australia}

\affiliation[c]{Institute for Theoretical Physics, Karlsruhe Institute of Technology\\
76128 Karlsruhe, Germany}

\affiliation[d]{Czech Technical University in Prague, FNSPE\\
Brehova 7, 115 19 Prague, Czech Republic}

\affiliation[e]{Institute of Nuclear Physics Polish Academy of Sciences\\
Radzikowskiego 152, 31-342 Kraków, Poland}

\emailAdd{miroslav.myska@fjfi.cvut.cz}

\abstract{Herwig 7 is a general-purpose Monte Carlo generator of particle collisions comprising both hard perturbative as well as soft phenomenological physics. Herwig is therefore capable to describe the entire final state of hadronized particles in a collision event. A spacetime topology of a parton system entering hadronization is fully described and tested for the first time. A combination of information from particles momenta and spacetime positions is utilized to minimize a boost-invariant distance measure of the parton system. We present a reasonable agreement of the model with a selection of experimental data and conclude that spacetime event topology can be meaningfully used in the further development.}

\FullConference{%
  40th International Conference on High Energy physics - ICHEP2020\\
  July 28 - August 6, 2020\\
  Prague, Czech Republic (virtual meeting)
}

\begin{document}
\maketitle

\section{Introduction}

The entire hadron event simulation in Herwig 7 \cite{Bahr:2008pv, Bellm:2017bvx, Bellm:2015jjp} is a highly complex task. Here we focus only
on the most relevant parts, which are unavoidably connected to the color reconnection procedure. It browses parton final state inside a particle collision event and searches for a place of further improvement of a color flow description. It can be done via pseudo-addition of an extra gluons interconnecting quarks entering hadronization and thus change their color charge. Such a mechanism is motivated by a need of corrections for errors in the leading-color approximation of the parton shower and for uncertainties in the correlations among multiple parton interactions (MPI) within one hadron collision.

Herwig 7 first focused on reconnecting excited $q\bar{q}$ pairs called
clusters (mesonic clusters now) by minimizing the sum of the invariant masses \cite{Gieseke:2012ft}. It was followed by \cite{Bierlich:2015rha},
where the possibility of creation of larger $qqq$ and $\bar{q}\bar{q}\bar{q}$ clusters (baryonic clusters) was introduced.
A list of different approaches involves color reconnection at the perturbative stages of event simulation or reconnection inspired by perturbative techniques \cite{Lonnblad:1995yk,Gieseke:2018gff,Bellm:2018wwz}. A deeper introduction of the topic goes beyond this proceedings and we refer reader to e.q. \cite{Sjostrand:2013cya}.

In the following lines we offer a short summary of the spacetime color reconnection model \cite{Bellm:2019wrh}, which is based on baryonic reconnection but utilizes also a newly available information about positions of all final state partons in the spacetime.

\section{Event generation and spacetime positions}

The MPI model distinguishes hard and soft interactions with a tunable value of the cut $p_{\perp}^{\mathrm{min}}$. The hard interactions are calculated perturbatively and initial-and final-state partons undergo showering.
Their color topology is motivated by the leading-color approximation used in the shower.
The soft scatterings are modeled phenomenologicaly according to multiperipheral model \cite{Bahr:2009ek,Gieseke:2016fpz}. Both types of interactions are bound together through the optical theorem \cite{Borozan:2002fk}, which leads to the calculation of the total cross section, $\sigma_{\mathrm{tot}}(s)$, of the given collision at a c.m.s energy squared ($s$), which is here taken also as a tunable parameter. We shortly note, that a new parameter $R_{\mathrm{Diff}}$ was used to describe the amplitude of the non-diffractive cross section, but which is not a regular part of Herwig 7. It turns out that the average number of MPI, $\langle n(b,s) \rangle$, at a given impact parameter of a collision, $b$, can be described using an hadron-hadron overlap function $A(b;\mu)$ and inclusive cross section $\sigma^{\mathrm{inc}}(s;p^{\mathrm{min}}_{\perp})$
\begin{equation}
A(b;\mu) \sigma^{\mathrm{inc}}(s;p^{\mathrm{min}}_{\perp}) = \langle n(b,s) \rangle.
 \label{eq:H7:eikonal}
\end{equation}
The prescription works for both hard and soft MPI, only with a different value of a parameter $\mu$, which has a meaning of inverse proton radius squared. The hard one, $\mu_{\mathrm{hard}}$, is a tunable parameter, while the soft one, $\mu_{\mathrm{soft}}$, 
can be determined with respect to other parameters. In this work, overlap function is chosen to be a Bessel function of the third kind:
\begin{equation}
  \label{eq:H7:overlap}
  A(b;\mu) = \frac{\mu^2}{96\pi}(\mu b)^3K_3(\mu b).
\end{equation}
Given the mean number of MPI, we can now randomly sample the actual number of hard and soft
interactions using a Poissonian distributions. Newly, we are also able to sample the impact parameter $b$ for each event and, thus, we can assign the position of proton centers in the transverse space. The overlap function then governs the density of MPI scattering centers
and we can randomly sample the position of each of them in the transverse plane.

As already mentioned, partons associated with the hard interactions are showered, meaning that they can undergo splitting and further propagation in spacetime according to their actual virtuality. We have found that the distance traveled by a parton from his scattering center in the transverse plane is negligible except the final propagation at minimal virtuality. Therefore we have decided to further consider only this last propagation in the shower plus we assign the same value of the minimal virtuality  $\nu^2$ to all partons.  
The distance traveled by a parton results from a Lorentz transformation, where the parton's proper lifetime, $t^*$, is sampled from the exponential decay law
\begin{equation}
    P_{\text{decay}}(t < t^*) = 1 - \exp\left(-\frac{t^*}{\tau_{0}}\right),
    \label{eq:H7:decaylaw}
\end{equation}
where 
\begin{equation}
    \tau_{0} = \frac{\hbar m}{\nu^2}
    \label{eq:H7:approxLifetime}
\end{equation}
depends directly on parton's mass $m$ and our new tunable parameter $\nu^2$.
The last propagation through shower algorithm provides a spacetime position to all partons entering hadronization with respect to their scattering center. The final position is therefore a result of a combination of both sources of displacement: MPI position and shower propagation.

\section{Spacetime color reconnection}

Herwig 7 uses the cluster hadronization model \cite{WEBBER1984492}, based on the pre-confinement property of
angular-ordered showers \cite{AMATI197987}. The algorithm deals only with quarks (q) and antiquarks ($\bar{q}$) and combines them into a color-less $q\bar{q}$ pairs called mesonic clusters. All hadronizing gluons have to be first non-perturbatively split to quarks.
Because of the chosen leading-color topology for the additional scatters,
MPI are color-connected to the beam remnant and other subprocesses. It
is shown in \cite{Gieseke:2012ft} that this approximation performs significantly worse for the soft MPI and thus an extra treatment is required. The clusters undergo
the color reconnection. It aims to minimize a chosen measure, typically
momentum-based, of the given set of clusters. Herwig 7 has currently three
available models \cite{Gieseke:2012ft,Gieseke:2017clv}: Plain, Statistical/Metropolis, and Baryonic.

Here, we briefly describe the main principle of the new baryonic spacetime model of color reconnection. The original model \cite{Gieseke:2017clv} was modified to incorporate spacetime separation among partons. It starts with the preselection of mesonic clusters, which are found to be suitable for the reconnection. It is done by 
a loop of checks, where a relative rapidity is calculated for quarks and for antiquarks in the frame of the first cluster’s quark axis. At the end of a sub-loop for one cluster, three possible recommendations can come forth: do nothing, do mesonic reconnection, do baryonic reconnection. 

The second case, i.e. swapping antiquarks within two mesonic cluster system, happens with a flat probability if this criterion is satisfied:
\begin{equation}
    R_{q\bar{q}'} + R_{q'\bar{q}} < R_{q\bar{q}} + R_{q'\bar{q}'}.
    \label{eq:H7:MesoRecoCriterion}
\end{equation}
Here, $q'$ and $\bar{q}'$ are partons from the second cluster and $R$ is our new measure
\begin{equation}
    R_{ij}^2 = \frac{\Delta r_{ij}^2}{d_0^2} + \Delta y_{ij}^2 ,
    \label{eq:H7:spacetimeMeasure}
\end{equation}
\noindent which can be calculated generally for any partons $i,j$.
This measure contains a combination of spacetime transverse distance 
between the constituents
$\Delta r_{ij}^2 = (\vec{x}_{\perp,i} - \vec{x}_{\perp,j})^2 $
and rapidity difference $\Delta y_{ij}$.
$d_0$ is the characteristic length scale for color reconnection
in our spacetime model, which is a tunable parameter.
This parameter describes the strength of 
the transverse component of the spacetime measure
relative to the rapidity component.

The third case, i.e. creating two baryonic clusters out of three mesonic clusters, is done with a tunable probability $w_b$ if
\begin{equation}
    R_{q,qq} + R_{\bar{q},\bar{q}\bar{q}} < R_{q,\bar{q}} + R_{qq,\bar{q}\bar{q}},
\end{equation}
where we first find a pair of quarks (and antiquarks) with the lowest possible measure \eqref{eq:H7:spacetimeMeasure}.

In summary, the new model still relies on the preselection of clusters based on relative rapidity, however the transverse separation between constituents in measure \eqref{eq:H7:spacetimeMeasure} provides additional information and thus improves the
original baryonic color reconnection model, especially in larger systems like heavy-ion collisions, as we forsee. We would like to note that the building of 
baryonic spacetime color reconnection model in Herwig 7 required additional changes 
to the original code and we refer reader to a separate publication \cite{Bellm:2019icn}.

\section{Results}

At the end, we put the modified version of Herwig 7 to the test and tune its most relevant parameters
to best describe the experimental data for proton-proton collisions at 7 TeV \cite{Adam:2015qaa,Abelev:2012sea,Aad:2010ac,Aad:2010fh,Khachatryan:2015gka}.

\begin{table}[h!]
\centering
\begin{tabular}{lcccccc}
      $\sigma_{\mathrm{tot}}~[\mathrm{mb}]$ & $R_{\mathrm{Diff}}$&
      $p^{\mathrm{min}}_{\perp}~[\mathrm{GeV}]$&$\mu^2_{\mathrm{hard}}~[\mathrm{GeV}^2]$&$\nu^2~[\mathrm{GeV}^2]$ & $d_0~[\mathrm{fm}]$ & $w_b$ \\
\hline
 $96.0$ & $0.2$  & $3.0$  & $1.5$ & $4.5$ & $0.15$ & $0.98$ \\
\end{tabular}
  \caption{The best-fit values for the Minimum Bias model parameters including the new spacetime color reconnection.}
  \label{tab:H7:tune}
\end{table}

\begin{figure}[thb]
\centering
\includegraphics[width=0.45\textwidth]{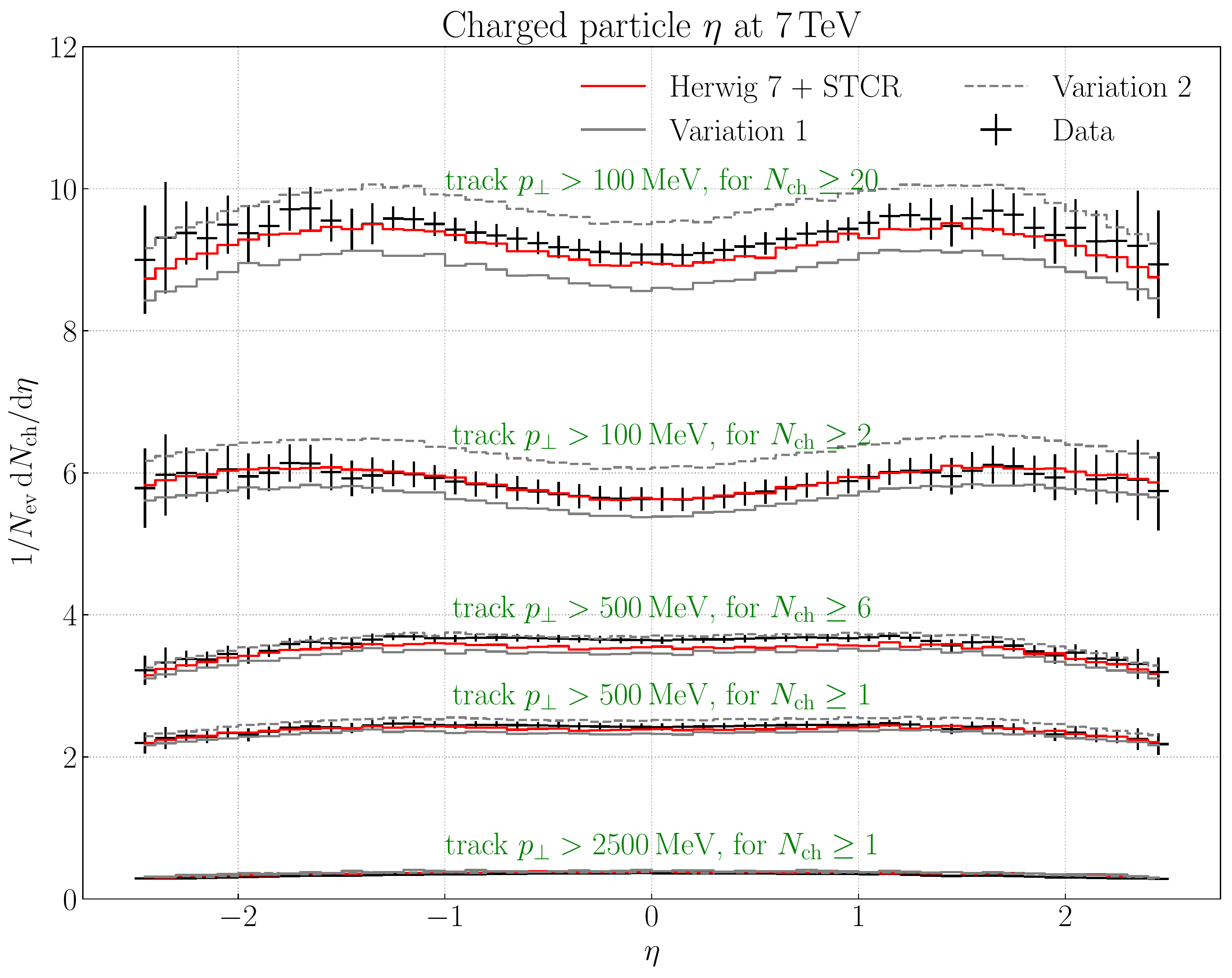}
\includegraphics[width=0.45\textwidth]{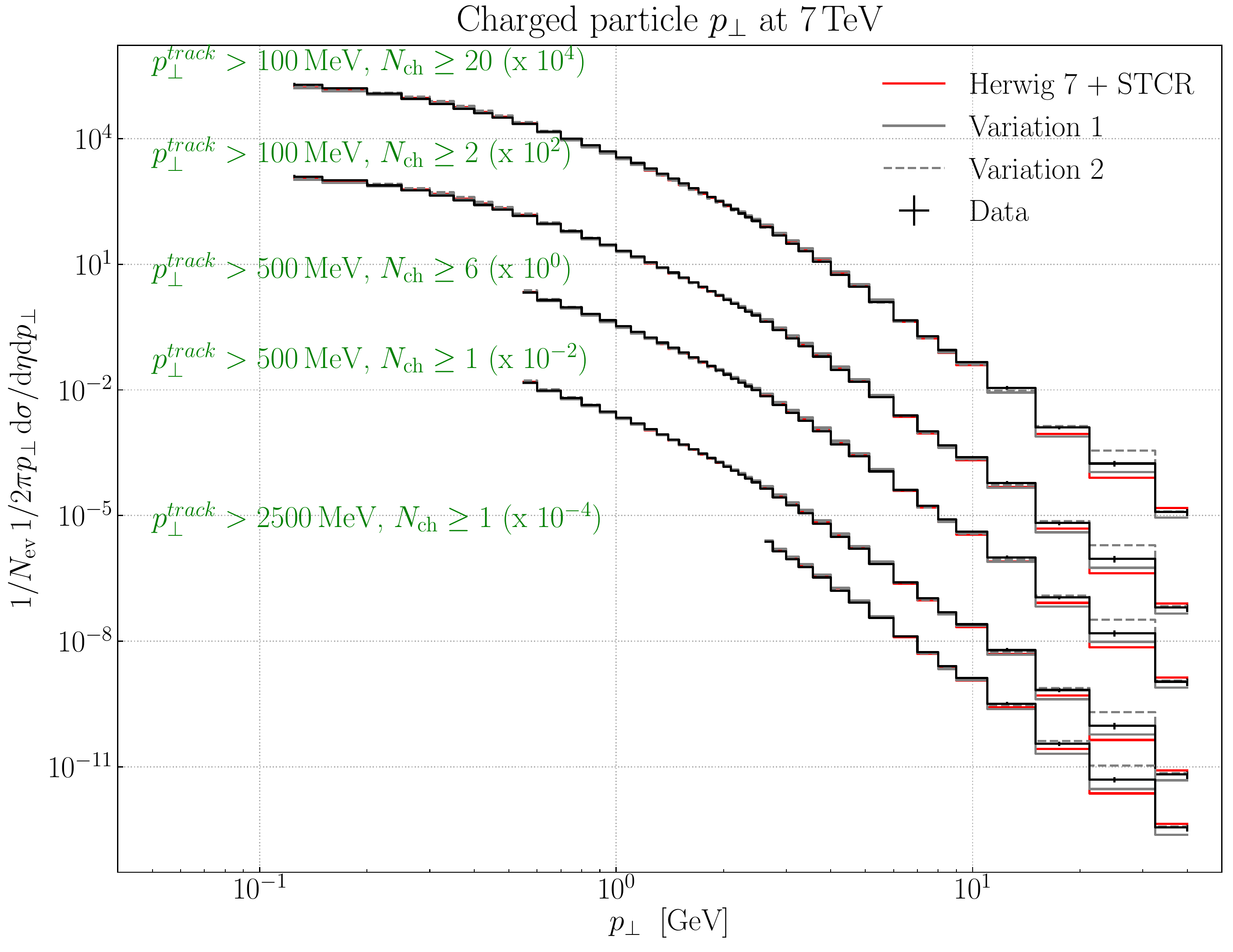}
\caption{Normalized distributions for charged particls against rapidity (left) and transverse
momentum (right) for various leading track $p_{\perp}$ and number of charged particles $N_{\mathrm{ch}}$. 
The displayed variations are purely in the spacetime length and minimum virtuality parameters of our model.}
\label{fig:H7:result}
\end{figure}

The final results are summarized in Table \ref{tab:H7:tune} and correspond to the red line in Figure \ref{fig:H7:result} labeled as Herwig 7 + STCR.
The list of parameters of interest include the three parameters of our spacetime baryonic color reconnection ($\nu^2$, $d_0$, $w_b$)
and four parameters necessary to well describe MPI
($\sigma_{\mathrm{tot}}(s)$, $R_{\mathrm{Diff}}$, $\mu^2$, $p^{\mathrm{min}}_{\perp}$), which were described above.
The tuning was performed using the Autotunes \cite{Bellm:2019owc}, which is build on the Rivet \cite{Buckley:2010ar} and Professor \cite{Buckley:2009bj} frameworks. On top of the best fit, we also show two variations for different values of $\nu^2$ and $d_0$ parameters to demonstrate the variability of the color reconnection model, see Figure \ref{fig:H7:result}. For the parameter $w_b$, we would like to emphasize that the tuned value of 0.98 might seem too close to unity. It is given by the fact
that the color reconnection algorithm first pre-selects clusters (using the relative rapidity span) for spacetime measure check.

Figure \ref{fig:H7:result} shows the rapidity and transverse momentum distributions as measured in \cite{Aad:2010ac}, where we can observe very good agreement between model and data. We can clearly declare that the model of color reconnection based on the spacetime topology of parton-level event is capable to describe Minimum Bias event measurements.
Not every possible observable is of course described that well as shown here, but it is a general feature of Herwig 7 and not of the presented model. More results and discussion can be found in \cite{Bellm:2019wrh}.

\section*{Acknowledgments}

This work has received funding from the European Union’s Horizon 2020 research and innovation programme 
as part of the Marie Sklodowska-Curie Innovative Training Network MCnetITN3 
(grant agreement no. 722104). JB acknowledges funding by the European Research 
Council (ERC) under the European Union's Horizon 2020 research and innovation programme, grant agreement No 668679. 
This work has been supported by the BMBF under grant number 05H18VKCC1.
CBD is supported by the Australian Government Research Training 
Program Scholarship and the J. L. William Scholarship. AS acknowledges support from the National Science Centre, Poland Grant No. 2016/23/D/ST2/02605. MM acknowledges the support by the grant 18-07846Y of the Czech Science
Foundation (GACR).

\end{document}